\documentclass[aps,prd,reprint,preprintnumbers,superscriptaddress,showpacs,twocolumn]{revtex4-1}
\usepackage{latexsym,graphicx,amssymb,amsmath,mathrsfs}
\usepackage{setspace,bm}
\usepackage[breaklinks, colorlinks=true, pdfstartview=FitV, linkcolor=red, citecolor=blue, urlcolor=blue]{hyperref}
\usepackage[usenames]{color}
\usepackage{latexsym}
\usepackage{epstopdf}
\usepackage{mathtools}
\usepackage{url}
\usepackage{comment}
\usepackage{braket}
\usepackage{bbm}
\usepackage{CJKutf8}
\usepackage{ulem}
\usepackage{xcolor}
\DeclareRobustCommand{\erase}{\bgroup\markoverwith{\textcolor{red}{\rule[.5ex]{2pt}{0.4pt}}}\ULon}

\begin{document}

\title{Path optimization method for the sign problem caused by fermion determinant}

\author{Kazuki Hisayoshi}
\affiliation{Department of Computer Science and Engineering, Faculty of Information Engineering, Fukuoka Institute of Technology, Fukuoka 811-0295, Japan}

\author{Kouji Kashiwa}
\email[]{kashiwa@fit.ac.jp}
\affiliation{Department of Computer Science and Engineering, Faculty of Information Engineering, Fukuoka Institute of Technology, Fukuoka 811-0295, Japan}

\author{Yusuke Namekawa}
\affiliation{Education and Research Center for Artificial Intelligence and Data Innovation, Hiroshima University, Hiroshima 730-0053, Japan}

\author{Hayato Takase}
\noaffiliation{}

\begin{abstract}
The path optimization method with machine learning is applied to the one-dimensional massive lattice Thirring model, which has the sign problem caused by the fermion determinant.
This study aims to investigate how the path optimization method works for the sign problem. 
We show that the path optimization method successfully reduces statistical errors and reproduces the analytic results.
We also examine an approximation of the Jacobian calculation in the learning process and show that it gives consistent results with those without an approximation.
\end{abstract}
\maketitle

\section{Introduction}
\label{sec:introduction}

To understand several important properties of quantum chromodynamics (QCD), such as the chiral phase transition and the confinement-deconfinement transition, the Markov chain Monte Carlo (MCMC) method is an important tool.
In the MCMC calculation with the Boltzmann weight, expectation values are estimated using an action for the probability distribution.
However, the effective action can become complex at finite chemical potential even if the partition function itself is real.
The weight can no longer be considered as a probability.
Reweighting the imaginary part of the action is possible, but inefficient, especially for large volumes.
This problem is called the sign problem; see Refs.\,\cite{deForcrand:2010ys,Alexandru:2020wrj,Nagata:2021bru,*Nagata:2021ugx}.

Since the partition function has an integral representation, we can address the sign problem by optimizing the integration path on the complexified dynamical variable plane with such as the path optimization method (POM)~\cite{Mori:2017pne,Mori:2017nwj} or the sign optimized manifold~\cite{Alexandru:2018fqp}, which is related to the Lefschetz thimble method~\cite{Witten:2010cx,Cristoforetti:2012su,Fujii:2013sra} and the convex optimization~\cite{Lawrence:2023sfc}.
The modification of the integration path does not change the integral as long as Cauchy's integral theorem holds.
The path optimization method has been applied to several models~\cite{Mori:2017nwj,Kashiwa:2018vxr,Bursa:2018ykf,Kashiwa:2019lkv,Mori:2019tux,Kashiwa:2020brj,Namekawa:2021nzu,Namekawa:2022liz,Giordano:2022miv,Rodekamp:2022xpf,Rodekamp:2023byu,Kanwar:2023otc,Lin:2023svo}, and measurement of observables~\cite{Detmold:2020ncp,Detmold:2021ulb,Bedaque:2023ovz}.
In particular, the authors investigated the $0+1$ dimensional QCD in which the sign problem is induced by the quark determinant term~\cite{Mori:2019tux}.
It is important to extend the calculation to four-dimension, but the numerical cost is still very high.
It motivates us to test the path optimization with machine learning for the one-dimensional Thirring model~\cite{THIRRING195891} as a laboratory, which has the same origin of the sign problem as that of QCD.
The Thirring model has been studied using the Lefschetz thimble method and its extensions~\cite{Fujii:2015bua,Fujii:2015vha,Alexandru:2015xva,Alexandru:2015sua,Fukuma:2017fjq,DiRenzo:2021kcw}, the sign-optimized manifold approach~\cite{Alexandru:2018fqp,Alexandru:2018ddf}, and the subtraction method~\cite{Lawrence:2022dba}.
It is an important check if the path optimization method using machine learning can reduce the sign problem as the other methods.
We note that an alternative model which has close relation in terms of the sign problem in QCD is the Stephanov model~\cite{Stephanov:1996ki,Halasz:1998qr}. 
The Stephanov model is one of the matrix models corresponding to QCD in the large $N$ limit, where $N$ is the number of colors.
The complex Langevin method was applied, but the analytic results could not be reproduced~\cite{Bloch:2017sex}.
In contrast, the worldvolume Hybrid Monte Carlo method can reduce the sign problem and reproduce the analytic results~\cite{Fukuma:2020fez}.
The path optimization using simple contour deformation was also applied, which significantly reduced the sign problem~\cite{Giordano:2023ppk}.

The path optimization method was first proposed in Ref.\,\cite{Mori:2017pne}, but machine leaning had not yet been introduced. Later, machine learning was introduced to describe the modified integral path in Ref.\,\cite{Mori:2017nwj}. 
Machine learning was also introduced to learn the manifold of the generalized Lefschetz-thimble method in Ref.\,\cite{Alexandru:2017czx}, where the results of the generalized Lefschetz thimble method were used as training data.
This is the first paper which employs supervised learning to avoid the sign problem. 
The details are reviewed in Ref.\,\cite{Alexandru:2020wrj}.

We also investigate a Jacobian approximation in the learning process for the lattice Thirring model.
Since Jacobian calculation is dominant in the learning process, the cost reduction is highly desirable.
It is important to evaluate efficiency of the Jacobian approximation for the model in which the sign problem occurs from the fermion determinant.

This paper is organized as follows.
In Sec.~\ref{sec:formulation}, we explain the formulation of the one-dimensional massive lattice Thirring model and the path optimization method.
The numerical setup is shown in Sec.~\ref{sec:setup}.
The numerical results are shown in Sec.~\ref{sec:results}, and Sec.~\ref{sec:summary} is devoted to the summary.

\section{Formulation}
\label{sec:formulation}

We employ the one-dimensional massive lattice Thirring model~\cite{Pawlowski:2013pje,Fujii:2015vha} as a laboratory to investigate the sign problem caused by the fermion determinant term.
First, we explain the formulation of the Thirring model.
Next, we explain the application of the path optimization method with machine learning to the model.

\subsection{One-dimensional massive Thirring model}

The action of the one-dimensional lattice Thirring model with one flavor on a $L$ lattice is given by
\begin{align}
    S &= S_\mathrm{F} + S_\mathrm{B},
\end{align}
where the fermion and boson parts are~\cite{Pawlowski:2013pje,Fujii:2015vha},
\begin{align}
    S_\mathrm{F}
    & = - \sum_{n=1}^L
          \bar{\chi}_n \Bigl\{ e^{i\tilde{A}_n} \chi_{n+1} - e^{-i\tilde{A}_n} \chi_{n-1} 
        + ma \chi_n  \Bigr\},    
    \nonumber\\
    S_\mathrm{B} &= \beta \sum_{n=1}^L (1-\cos A_n),
\end{align}
here $\chi_n$ is fermion field at site $n$, $\tilde{A}_n$ denotes $A_n - i \mu a$, $A_n$ is a bosonic auxiliary field coupled to the vector current, $ma$ and $\mu a$ are the mass and chemical potential in the lattice unit $a$, respectively, and $\beta$ is the inverse coupling.
To make the auxiliary field $A_n$ compact, the cosine function is introduced in $S_\mathrm{B}$~\cite{Fujii:2015vha}.
We set $a=1$ in all our calculations.
We impose the antiperiodic boundary condition for the fermion field, and thus the system becomes thermal.

The partition function ${\cal Z}$ can be represented after integration of the fermion fields as
\begin{align}
    {\cal Z}(\beta,\mu)
    &= \int {\cal D}A {\cal D}\chi {\cal D} {\bar \chi} \, e^{-S}
    \nonumber\\
    &= \int {\cal D}A \, e^{- \beta \sum_{n=1}^L (1-\cos A_n)
                            + \log \mathrm{det} \, D[A] },
    \label{eq:partition_function}
\end{align}
where
\begin{align}
    \mathrm{det}\, D
    &= \frac{1}{2^{L-1}} \Bigl[ \cosh \Bigl(L \hat{\mu} + i \sum_{n=1}^{L} A_n \Bigr)
     + \cosh (L \hat{m}) \Bigr],
\end{align}
here $\hat{m} = \sinh^{-1} (ma)$ and $\hat{\mu} = \mu a$; for details of the model, see Ref.\,\cite{Fujii:2015vha} and references therein.
The field $A_n$ acts similarly to the gluon field in QCD, which leads to the appearance of the sign problem at finite $\mu \in \mathbb{R}$, although the structure is simpler.
In the case of $\mu=0$, there is no sign problem.

Analytic results of the fermion condensate $s$ and the number density $n$ of the model are known~\cite{Fujii:2015vha} as
\begin{align}
    s &= \langle {\bar \chi} \chi \rangle
      \nonumber\\
      &= \frac{I_0^L(\beta) \sinh (L\hat{m})}{I_1^L(\beta) \cosh(L\hat{\mu}) + I_0^L(\beta) \cosh(L\hat{m}) }
      \frac{1}{\cosh(\hat{m})},
\label{eq:fermion_condensate}
\end{align}
and
\begin{align}
    n &= - \frac{1}{L} \left\langle \frac{\partial S}{\partial \mu} \right\rangle
    \nonumber\\
      &= \frac{I_1^L(\beta) \sinh (L\hat{\mu})}{I_1^L(\beta) \cosh(L\hat{\mu}) + I_0^L(\beta) \cosh(L\hat{m}) },
\label{eq:number_density}
\end{align}
where $I_k(x)$ means the modified Bessel function of the first kind for $k = 0, 1$.
Since we have the analytic result, we can estimate the correctness of the path optimization method for the sign problem induced by the fermion determinant term.

\subsection{Path optimization method}

In the path optimization method, the dynamical variables, $v \in \mathbb{R}$, are complexified as
\begin{align}
    v \mapsto v' = v_\mathrm{R} + i v_\mathrm{I},
    \label{eq:modified_path}
\end{align}
where $v_\mathrm{R}, v_\mathrm{I} \in \mathbb{R}$.
Since we explain general structure of the neural network here, we introduce generic variables $v$; in the actual computations, $v_\mathrm{R}$ are corresponding to real parts of the integral variables.
We employ the artificial neural network~\cite{mcculloch1943logical,hebb2005organization,rosenblatt1958perceptron,hinton2006reducing} to represent the modified integral path as proposed in Ref.\,\cite{Mori:2017nwj}:
\begin{align}
    \underbrace{v_\mathrm{R}}_{\mathrm{input\,layer}} \to \mathrm{hidden~layer}
      \to \underbrace{v_\mathrm{I}}_{\mathrm{output\,layer}}.
\label{eq:layer}
\end{align}
The output layer is
\begin{align}
    v_{\mathrm{I} l} = v^{(F)}_l &= [ w_{lk}^{(F-1)} f(v_k^{F-1}) + b_l^{(F-1)}],
\end{align}
and the hidden layer is composed of
\begin{align}
    v_k^{(p+1)} &= w_{kj}^{(p)} v_j^{(p)} + b_k^{(p)},
\end{align}
where
\begin{align}
    v^{(p)}_j &=  w_{ji}^{(p-1)} f(v_i^{(p-1)}) + b_j^{(p-1)},
\end{align}
here $v^{(p)}$ means the quantities in the $p$-th hidden layer ($p=1,\cdots,F-1$) with $v^{(0)}=v_\mathrm{R}$, weight $W$ and bias $b$ are the parameters of the neural network.
The function $f(\cdot)$ denotes the activation function, and we set $f(\cdot) = \tanh(\cdot)$.
The $\tanh$ function can cause the vanishing gradient problem in deep neural networks.
However, since we do not use deep neural networks in this paper, the $\tanh$ function is acceptable.

There are several ways to reduce the cost of the Jacobian calculation, which is quite expensive in the naive computation:
the Gaussian or real approximations~\cite{Alexandru:2016lsn} that reduce ${\cal O} (N_\mathrm{d}^3)$ to ${\cal O}(N_\mathrm{d})$ where $N_\mathrm{d}$ means the total degree of freedom of considering theory/model, the Grady algorithm in the holomorphic gradient flow~\cite{Alexandru:2017lqr} that reduces ${\cal O} (N_\mathrm{d}^3)$ to ${\cal O}(N_\mathrm{d}^2)$, the diagonal ansatz of the Jacobian matrix~\cite{Alexandru:2018fqp} that reduces ${\cal O} (N_\mathrm{d}^3)$ to ${\cal O}(N_\mathrm{d})$, the nearest neighbor lattice site ansatz~\cite{Bursa:2018ykf} that reduces ${\cal O} (N_\mathrm{d}^3)$ to ${\cal O}(N_\mathrm{d})$, the worldvolume approach~\cite{Fukuma:2020fez} which requires only a combination of the Jacobian and the vector evaluated using the anti-holomorphic flow without the explicit form of the Jacobian in the MC update and thus the cost is reduced from ${\cal O} (N_\mathrm{d}^3)$ to $ {\cal O}(N_\mathrm{d}^{1-2})$, and the complex-valued affine coupling layers~\cite{Rodekamp:2022xpf} that reduce ${\cal O} (N_\mathrm{d}^3)$ to ${\cal O}(N_\mathrm{d})$.
In this study, we use a simple approximation that replaces the Jacobian with an identity matrix in the learning process~\cite{Namekawa:2022liz}.
Comparison with other methods is an important future work.

It should be noted that the above network represents a connected integration domain because it constructs a continuous function and then the real part of the integral variables are taken as the parametric variables; the neural network even with a single hidden layer can represent any kind of continuous function on a compact subset, provided that we have a sufficient number of units in the hidden layer, which is known as the universal approximation theorem~\cite{cybenko1989approximation,hornik1991approximation}.
There is a possibility that an infinite region of the integral can contribute the final results because we are using the Cauchy's integral theorem.
If we have a spreading tendency of configurations to the infinite region, we should care about the contributions, but we do not encounter the problem in this study.
We may additionally need the parallel tempering method \cite{swendsen1986replica,geyer1991markov,hukushima1996exchange} or the worldvolume approach~\cite{Fukuma:2020fez} to avoid the Ergodicity problem in the MC sampling, as we demonstrated in Ref.~\cite{Kashiwa:2020brj}.

To optimize the parameters of the neural network using the backpropagation method~\cite{rumelhart1986learning}, we need the cost function ${\cal F}$.
We use the following form
\begin{align}
    {\cal F}[w,b]
    &= \frac{1}{2} \int d v_\mathrm{R} \, |e^{i\theta(v_\mathrm{R})}-e^{i \theta_0}|^2 \, |J(v_\mathrm{R}) \, e^{-S(v')}|
    \nonumber\\
     &= | {\cal Z} | \Bigl[ \langle e^{i\theta} \rangle_\mathrm{pq}^{-1} - 1 \Bigr],
    \label{eq:cost_function}
\end{align}
with $\theta = \arg(e^{-S + \ln J})$ and $\langle e^{i\theta} \rangle_\mathrm{pq}$ is the so-called average phase factor (APF) where $\langle \cdots \rangle_\mathrm{pq}$ means the phase reweighted expectation values; see Ref.\,\cite{Mori:2017nwj} for details of the cost function.
The decrease in the cost function leads to the increase in APF.
We employ the exponential moving average for the cost function to stabilize the learning~\cite{Kashiwa:2020brj}.
$J(v_\mathrm{R})$ is Jacobian induced via the complexification, and $\theta_0$ means the phase of the partition function.
In this study, $\theta_0 = 0$ is manifested since the partition function is definitely real.
Based on the cost function, which reflects the seriousness of the sign problem, we can perform the training with configurations generated by using the Hybrid Monte Carlo (HMC) method~\cite{Duane:1987de}. 
Therefore, the path optimization method does not need labeled training data.
In this sense, the path optimization method is a self-supervised learning-like method.

Even after obtaining a good modified integral path, the Boltzmann weight is still complex, and thus the reweighting method~\cite{Ferrenberg:1988yz} is a possible choice.
In this work, we use the phase reweighting as
\begin{align}
    \langle {\cal O} \rangle
    &= \frac{\langle {\cal O} e^{i\theta} \rangle_\mathrm{pq}}
            {\langle e^{i\theta} \rangle_\mathrm{pq}},
\label{eq:pq}
\end{align}
where ${\cal O}$ represents observables.

\section{Numerical setup}
\label{sec:setup}

Our numerical codes are made using PyTorch~\cite{paszke2019pytorch}.
To evaluate the expectation values, we generate $1000$ configurations after thermalization using the HMC method.
The statistical error is estimated using the Jackknife method with a bin size of $50$; see Appendix \ref{sec:app_error} for details.

\begin{figure}
\centering
\includegraphics[keepaspectratio, scale=0.35]{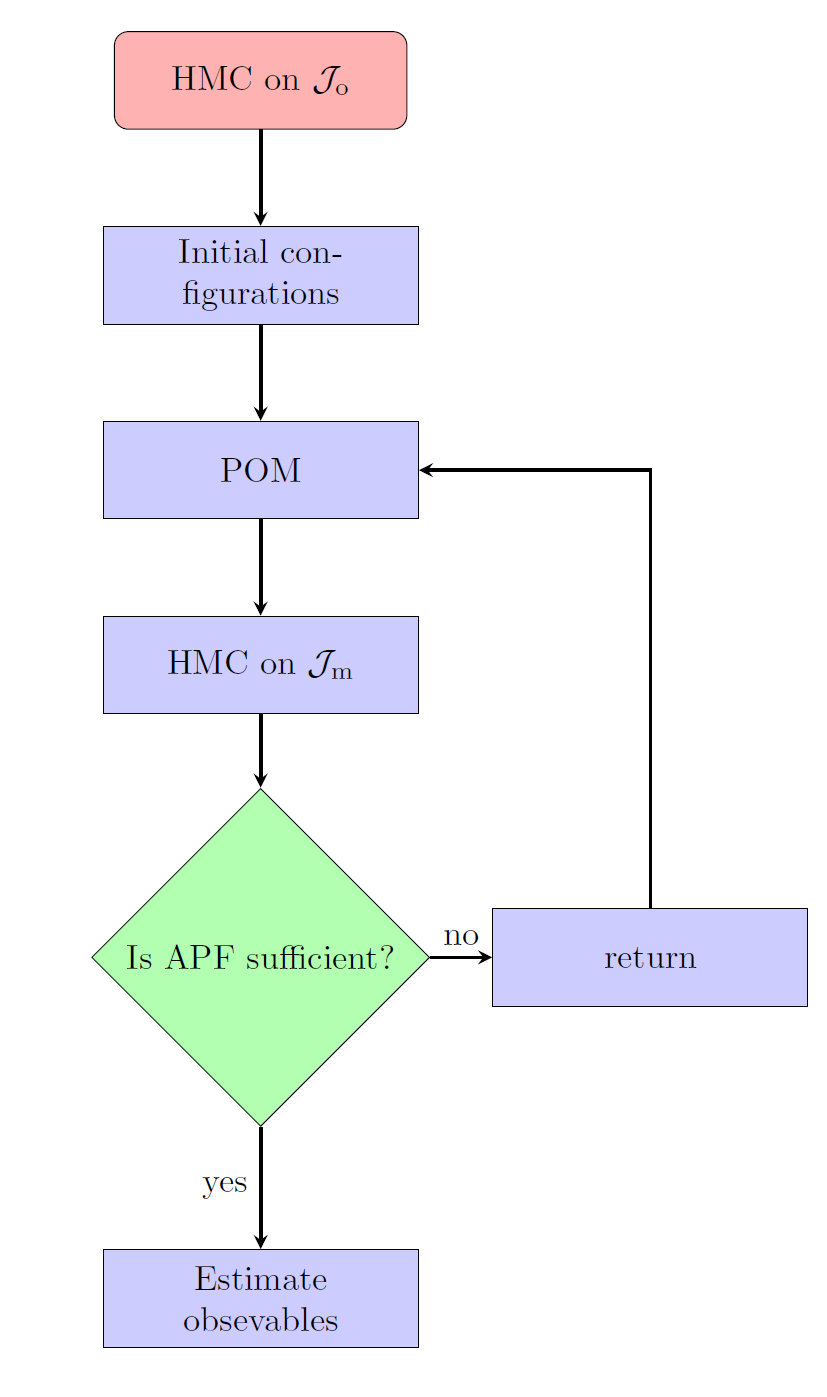}
\caption{The flowchart of the path optimization in this study. 
Symbols, ${\cal J}_\mathrm{o}$ and ${\cal J}_\mathrm{m}$, denote the original and the modified integral paths, respectively. 
The closed loop in the flowchart is one cycle (learning step) of the optimization procedure.}
\label{fig:flow_chart}
\end{figure}

Our procedure of the path optimization is presented in Fig.~\ref{fig:flow_chart}.
In the learning process, we employ AdamW~\cite{loshchilov2018decoupled}, one of the stochastic gradient methods, as an optimizer.
The number of units in the hidden layer is set to $64$.
We use a single hidden layer $F=1$, which is found to reduce the sign problem sufficiently in our setup.
Improvement of learning with the deep neural network is our future work.
In the one learning step, we use batch training~\cite{bottou1998online} with the batch size of $64$, and the parameters in the network are updated 30 times.

After training, we regenerate the configurations and estimate the average phase factor and observables.
This procedure is introduced to avoid the overtraining problem; we estimate the observables after the HMC update, not just after the training.
If the average phase factor is not sufficiently enhanced in the early stage of the training, we consider that the initial values of the neural network parameters are not good and thus restart the calculation with different initial values of parameters.
We stop learning in the $100$th step, which is enough to enhance the average phase factor.

\section{Numerical results}
\label{sec:results}

We show our numerical results for the one-dimensional Thirring model mainly with the lattice size $L=16$.
First, we show the full results, and later we argue an approximation of the Jacobian in the model.

\begin{figure}[b]
 \centering
 \includegraphics[keepaspectratio, scale=0.1]{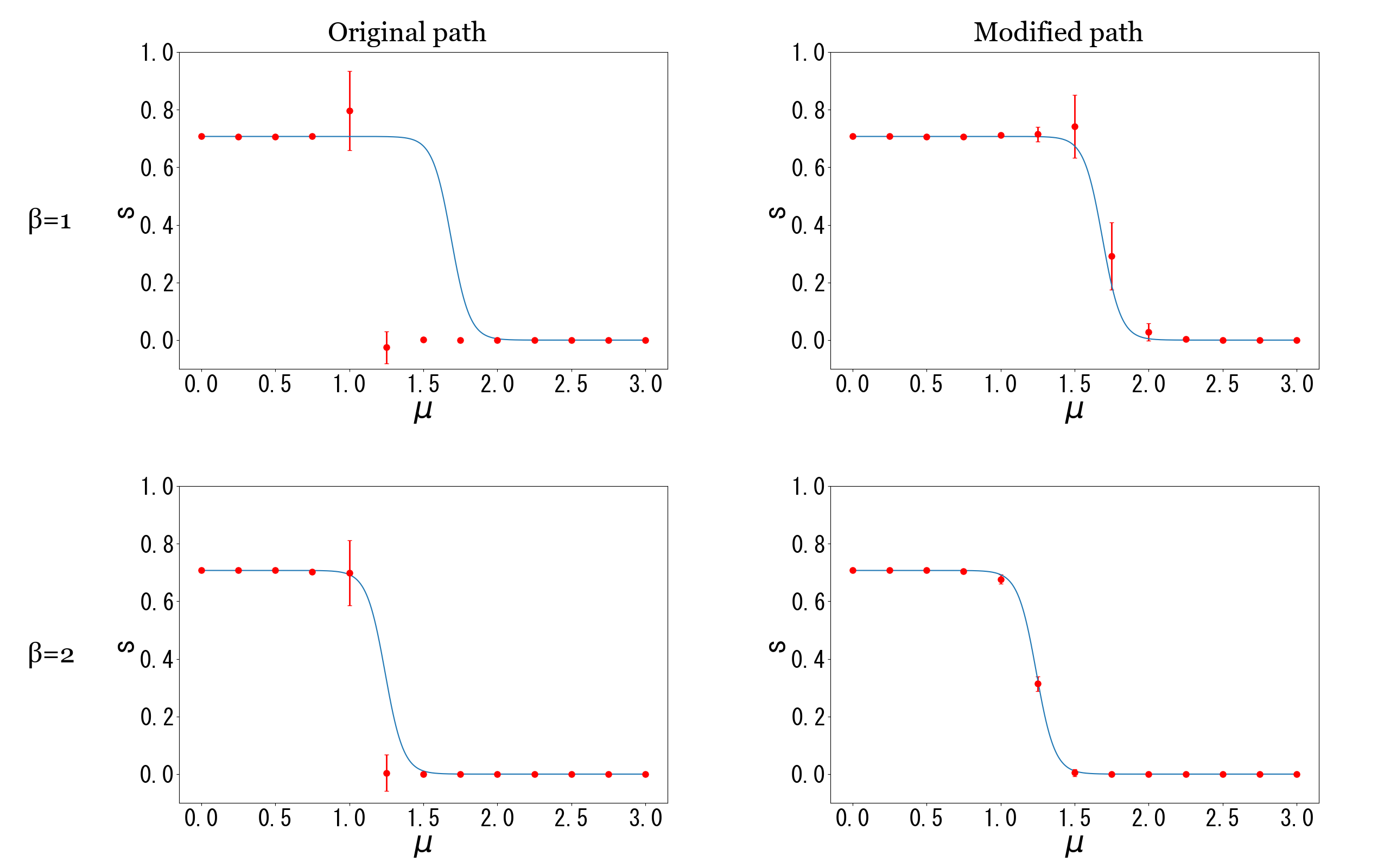}
 \caption{
 The $\mu$-dependence of the fermion condensate $s$ with $L = 16$.
 The symbols are our numerical results, and the lines denote the analytic results.
 The left (right) panels are the results on the original (modified) path.
 The upper (lower) panels are the results at $\beta =1$ ($\beta=2$).
 }
 \label{fig:mu_s}
\end{figure}

Figure~\ref{fig:mu_s} shows the $\mu$-dependence of the fermion condensate~\eqref{eq:fermion_condensate} at $\beta =1$ and $2$ on the original and modified paths.
\begin{figure}[t]
 \centering
 \includegraphics[keepaspectratio, scale=0.1]{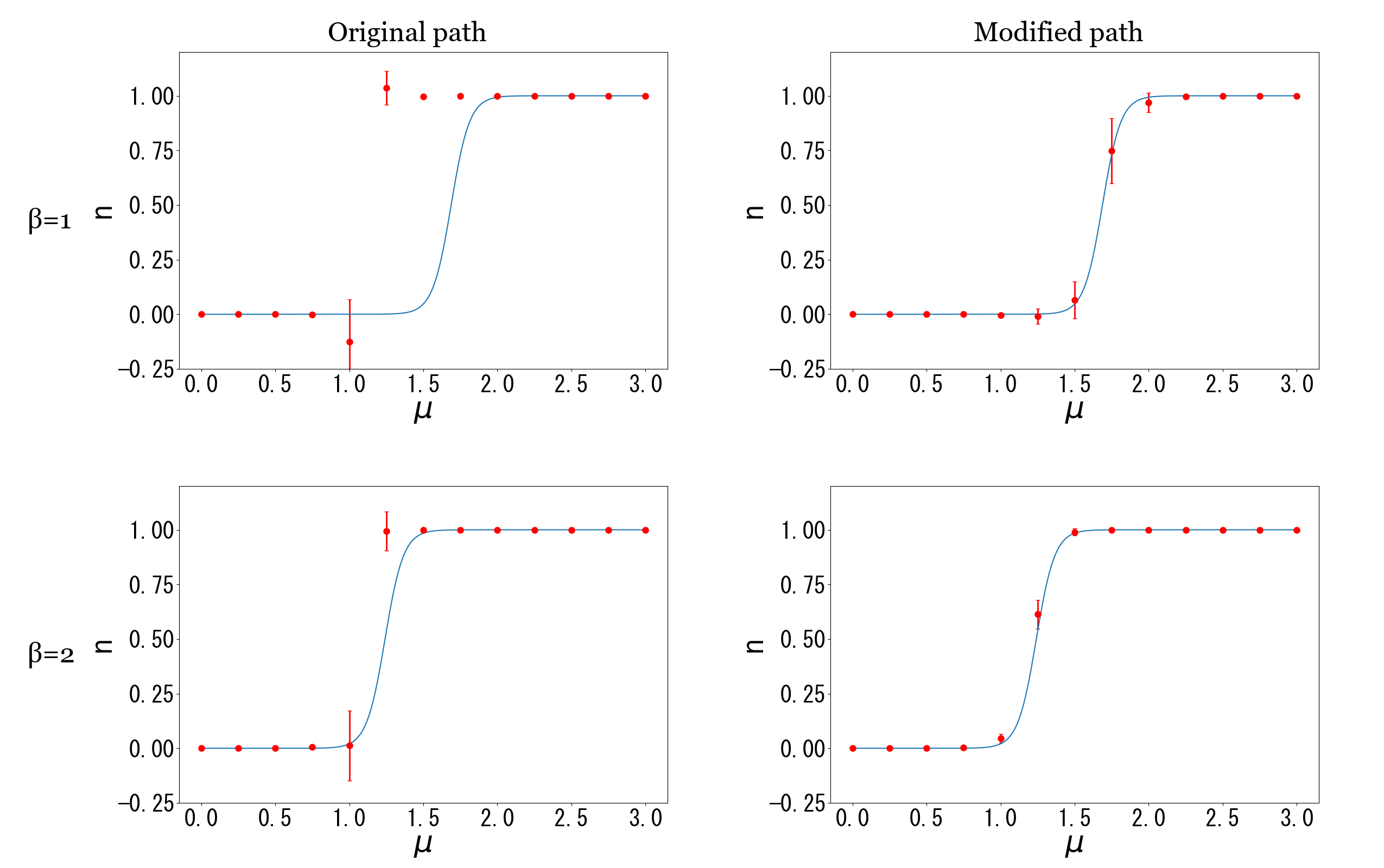}
 \caption{
 The $\mu$-dependence of the number density $n$ with $L = 16$.
 The symbols are our numerical results, and the lines denote the analytic results.
 The left (right) panels are the results on the original (modified) path.
 The upper (lower) panels are the results at $\beta =1$ ($\beta=2$).
 }
 \label{fig:mu_n}
\end{figure}
On the original path, the numerical results have huge errors due to the sign problem.
In addition, at $\beta=1$, the numerical results do not reproduce the analytical results with small errors.
This happens by an unbalanced sampling of configurations.
It will be relaxed if we increase the number of configurations.
On the modified integral path, we can reproduce the analytic results with small errors.
Figure~\ref{fig:mu_n} shows the $\mu$-dependence of the number density~\eqref{eq:number_density} at $\beta =1$ and $2$.
As in the case of the fermion condensate, our results on the modified path reproduce the analytic results with small errors, while those on the original path do not.

\begin{figure}[t]
 \centering
 \includegraphics[keepaspectratio, scale=0.1]{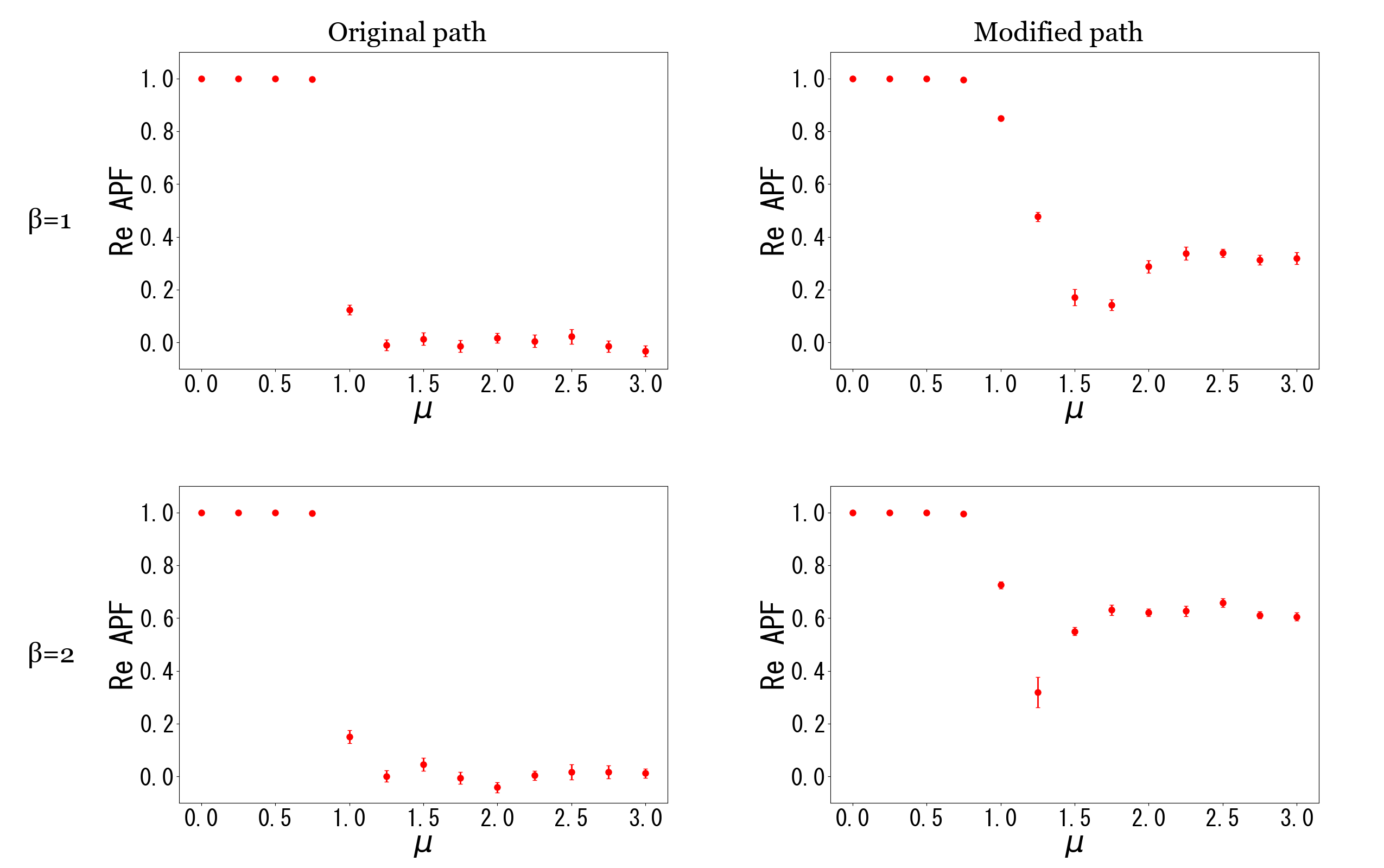}
 \caption{
 The $\mu$-dependence of the real part of APF with $L = 16$.
 The left (right) panels are the results on the original (modified) path.
 The upper (lower) panels are the results at $\beta =1$ ($\beta=2$).
 }
 \label{fig:mu_e}
\end{figure}

\begin{figure}[t]
 \centering
 \includegraphics[keepaspectratio, scale=0.2]{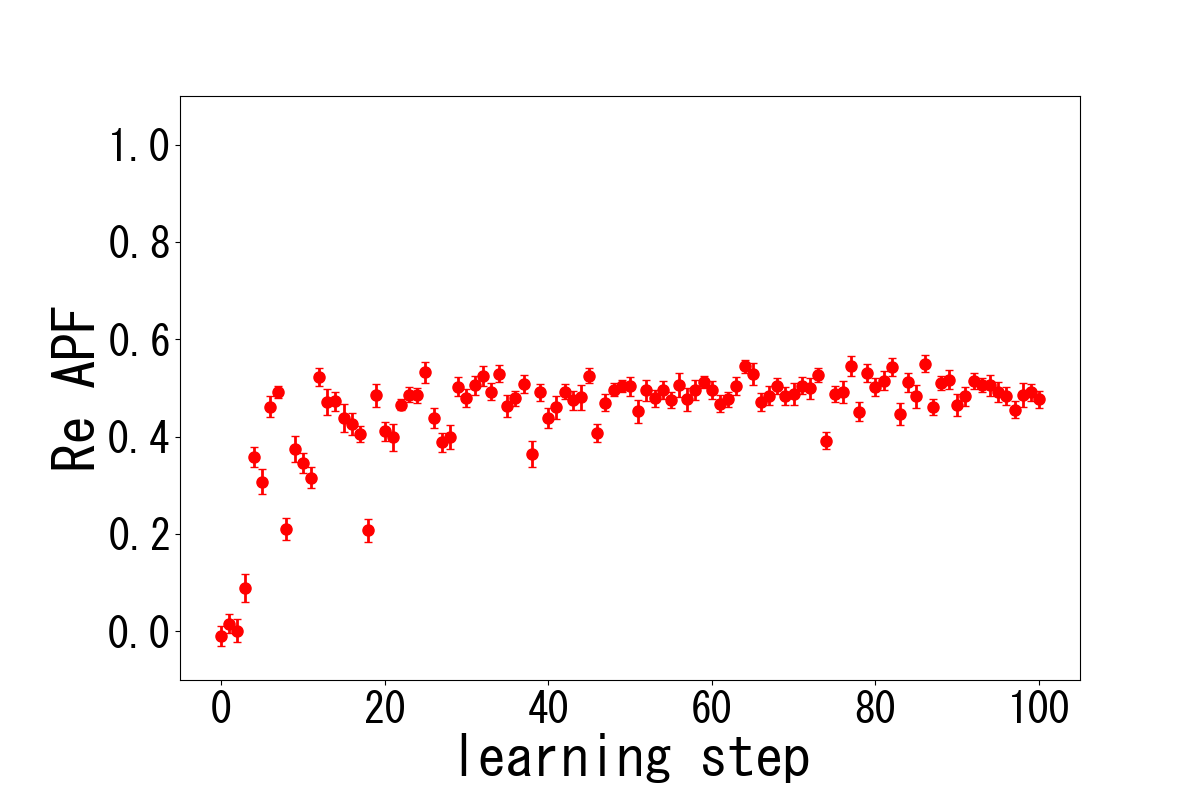}
 \caption{
 The learning step dependence of APF at $\beta=1$ and $\mu=1.25$ with $L = 16$.
 }
 \label{fig:APF_learning}
\end{figure}

Figure~\ref{fig:mu_e} shows the $\mu$-dependence of the real part of APF at $\beta =1$ and $2$.
Since the cost function used in this study is related to APF, we here show APF rather than the cost function itself.
The path optimization enhances APF, which leads to better-controlled errors.
Figure\,\ref{fig:APF_learning} shows how APF is improved by the path optimization; results at $\beta=1$ and $\mu=1.25$ are plotted as an example.
APF may be further improved by using a more complicated neural network because such a network has higher
expressive power; deep neural networks with the attention mechanism and networks which respect symmetry of the system as we employed in Ref.~\cite{Namekawa:2021nzu} could be a possible choice.

\begin{figure}[t]
 \centering
 \includegraphics[keepaspectratio, scale=0.07]{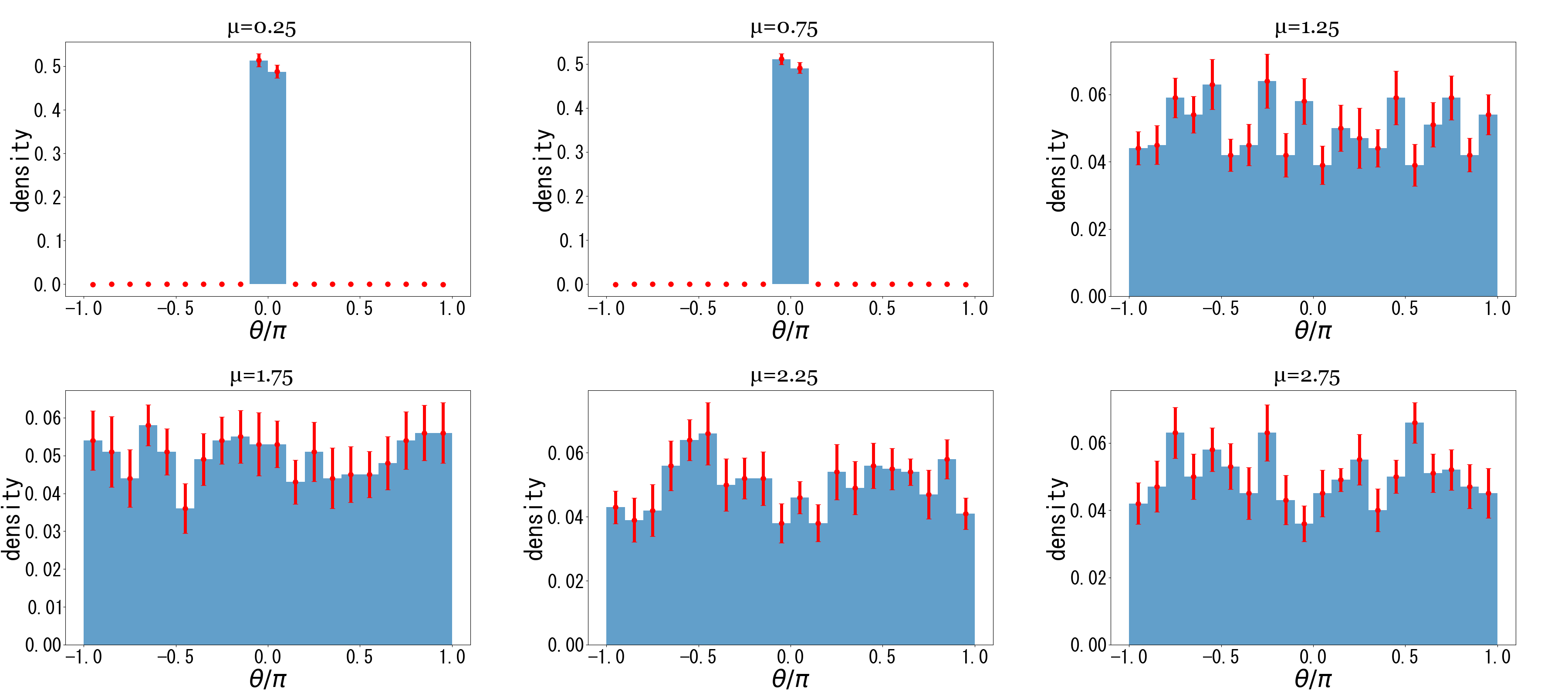}
 \caption{The histogram against $\theta$ on the original integral paths with $L=16$ at $\beta=1$.
 From the left-top to right-bottom panels, $\mu$ is set to $0.25$, $0.75$, $1.25$, $1.75$, $2.25$ and $2.75$.
 The histogram is normalized, and the statistical errors are also shown.
 }
 \label{fig:scatter_plot_APF_ori}
\end{figure}
\begin{figure}[b]
 \centering
 \includegraphics[keepaspectratio, scale=0.07]{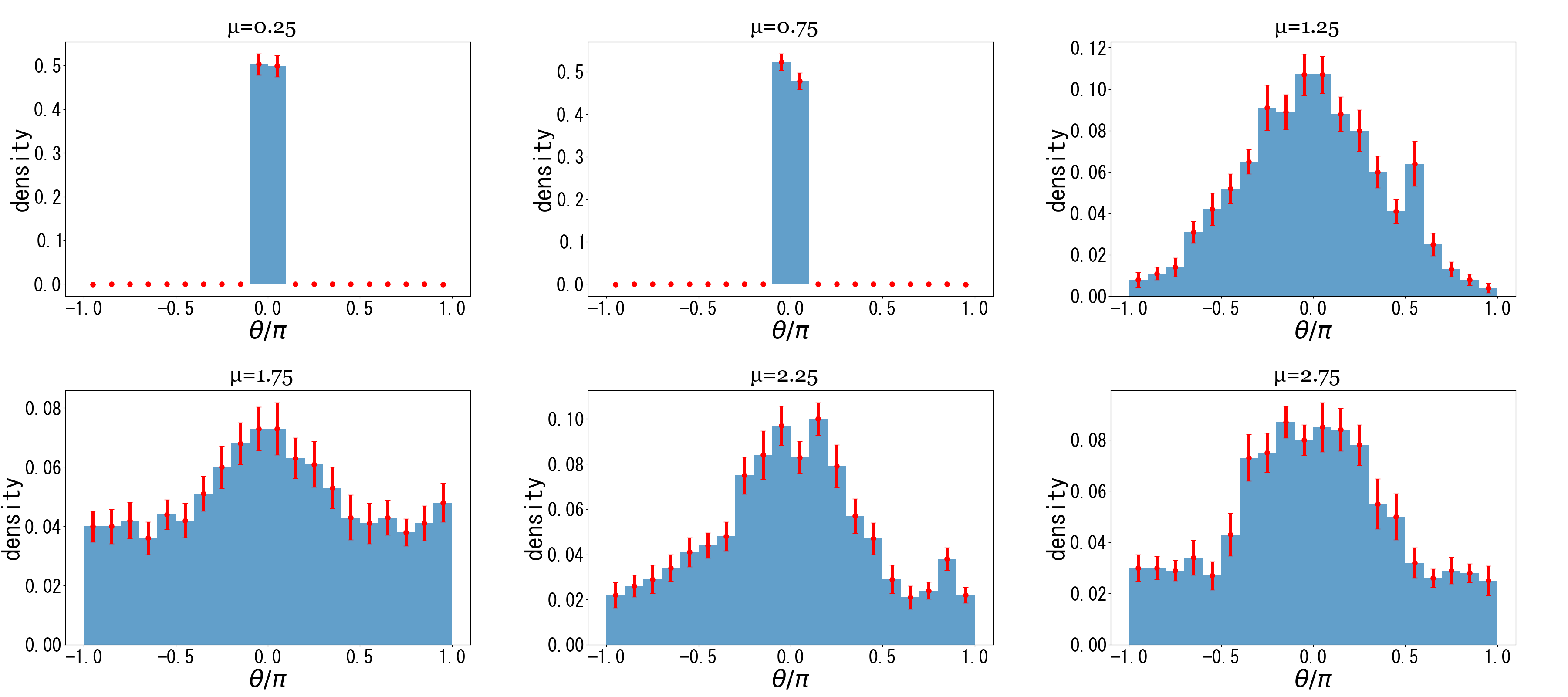}
 \caption{The histogram against $\theta$ on the modified integral paths with $L=16$ at $\beta=1$.
 From the left-top to right-bottom panels, $\mu$ is set to $0.25$, $0.75$, $1.25$, $1.75$, $2.25$ and $2.75$.
 The histogram is normalized, and the statistical errors are also shown.
 }
 \label{fig:scatter_plot_APF_trn}
\end{figure}
\begin{figure}[t]
 \centering
  \includegraphics[keepaspectratio, scale=0.07]{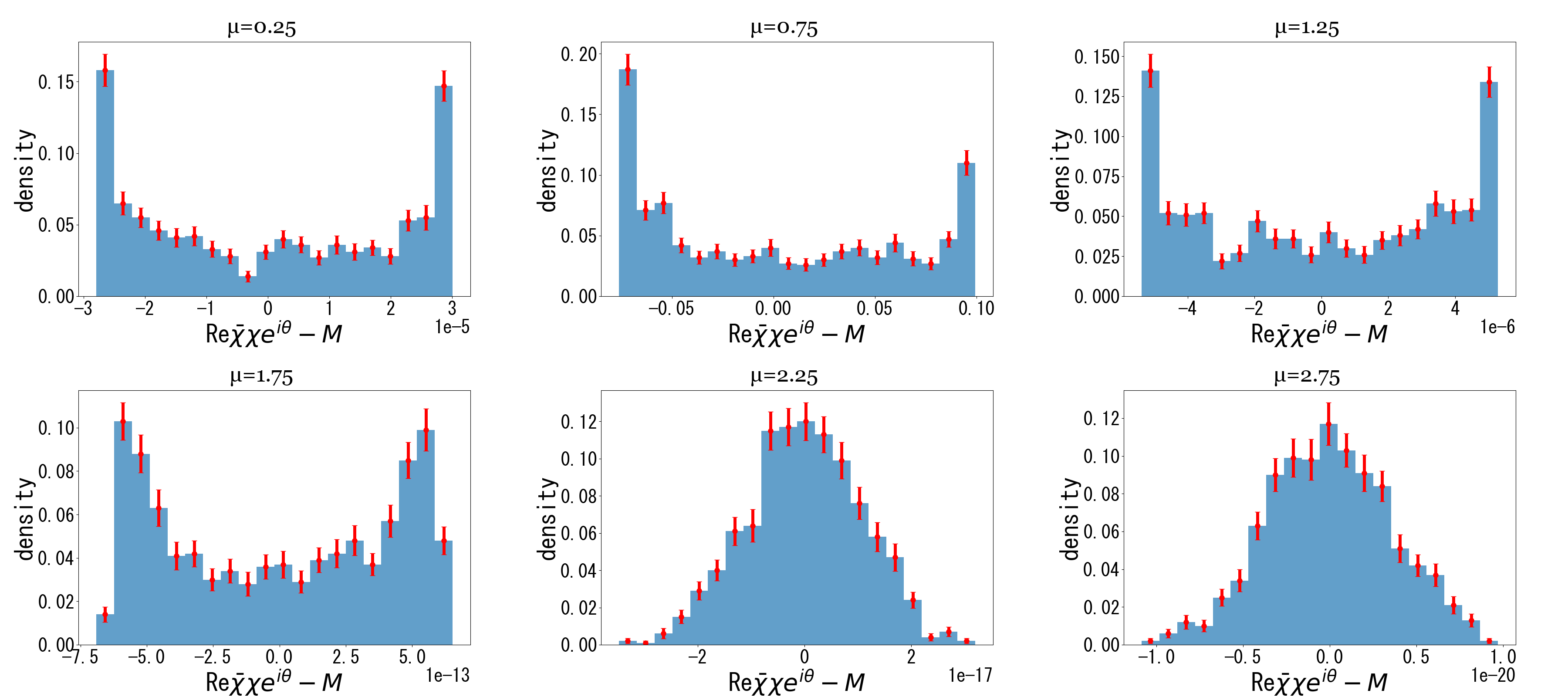}
 \caption{The histogram against $\mathrm{Re}\,\bar{\chi} \chi e^{i\theta} - M$ on the original integral paths with $L=16$ at $\beta=1$, where $M$ denotes the mean value of $\mathrm{Re}\,{\bar \chi} \chi e^{i\theta}$ at each $\mu$.
 From the left-top to right-bottom panels, $\mu$ is set to $0.25$, $0.75$, $1.25$, $1.75$, $2.25$ and $2.75$.
 The histogram is normalized, and the statistical errors are also shown.}
 \label{fig:scatter_plot_se_ori}
\end{figure}
\begin{figure}[b]
 \centering
 \includegraphics[keepaspectratio, scale=0.07]{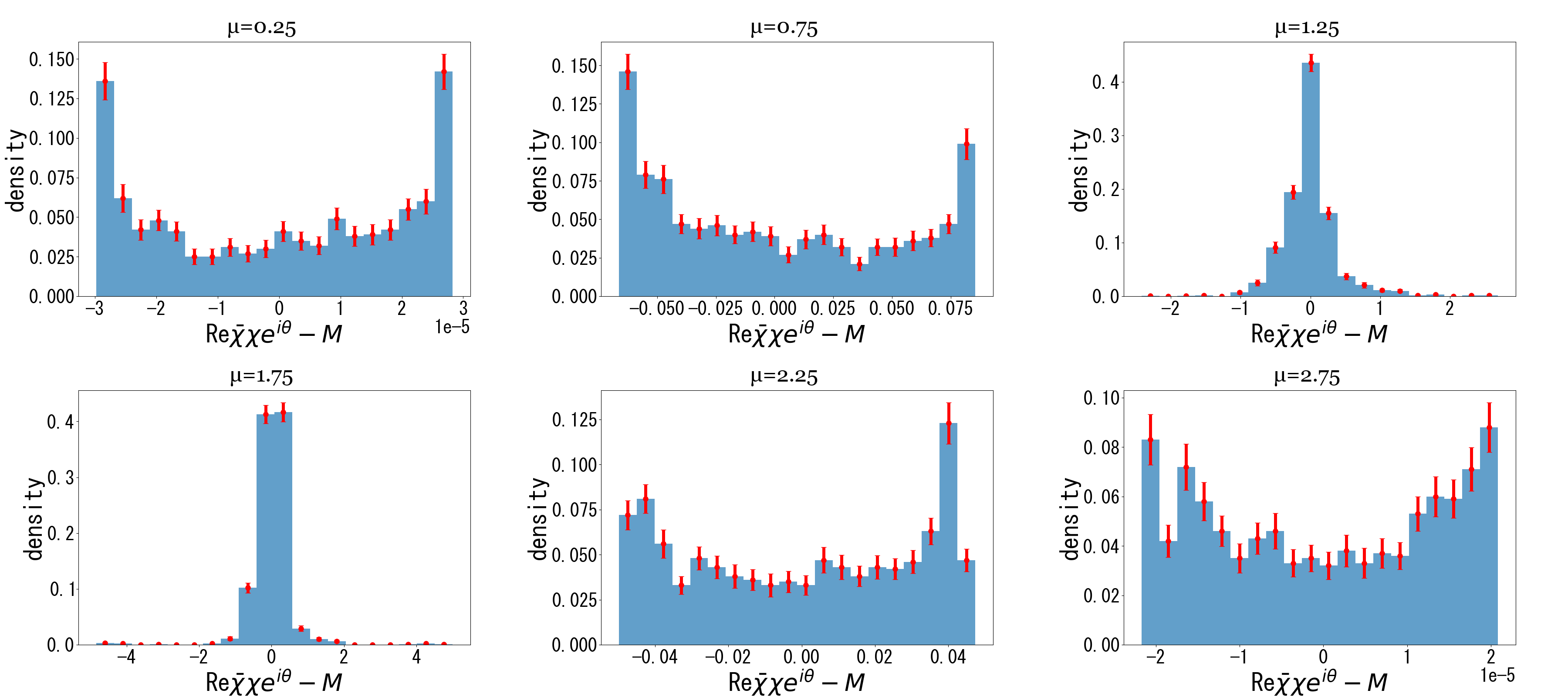}
 \caption{The histogram against $\mathrm{Re}\,\bar{\chi} \chi e^{i\theta}$ on the modified integral paths with $L=16$ at $\beta=1$, where $M$ denotes the mean value of $\mathrm{Re}\,{\bar \chi} \chi e^{i\theta}$ at each $\mu$.
 From the left-top to right-bottom panels, $\mu$ is set to $0.25$, $0.75$, $1.25$, $1.75$, $2.25$ and $2.75$.
 The histogram is normalized, and the statistical errors are also shown.
}
 \label{fig:scatter_plot_se_trn}
\end{figure}

The deformation of the integral path is visualized by histograms of the phase of APF on the original and deformed integral paths.
Below, we show normalized histograms at $\beta=1$ as an example.
The figures~\ref{fig:scatter_plot_APF_ori} and \ref{fig:scatter_plot_APF_trn} represent the histograms of the phase at $\beta =1$; the horizontal axis is normalized by $\pi$.
We can clearly see the difference between the histograms.
At $\mu = 0.25$ and $0.75$, the histogram is localized both on the original and modified paths, and AFP in Fig.~\ref{fig:mu_e} is close to one, indicating that the sign problem is mild.
At larger $\mu$, the histogram on the original path shows almost flat dependence on $\theta$, and AFP is close to zero, indicating that the sign problem is severe.
In contrast, the histogram on the modified path is still localized well, and AFP is non-zero, indicating that the sign problem is mild.
It is noted that at $\mu = 1.75$ we see a less clear peak in the histogram on the modified path.
It suggests that several thimbles contribute to the result.
As the Lefschetz thimble approach with parallel tempering (tempered Lefschetz thimble method)~\cite{Fukuma:2017fjq} and its extension to the continuous accumulation of deformed surfaces (worldvolume HMC method)~\cite{Fukuma:2020fez,Fukuma:2021aoo,Fukuma:2023eru} successfully evaluated contribution from many thimbles,
the path optimization combined with parallel tempering~\cite{Kashiwa:2020brj} may further improve the result.
This is our future work.
Figures~\ref{fig:scatter_plot_se_ori} and \ref{fig:scatter_plot_se_trn} exhibit histograms of $\mathrm{Re}\,\bar{\chi} \chi e^{i\theta}$ on the original and modified integral paths at $\beta = 1$; the subtraction constant $M$, which is explained in the figure caption, is introduced to show the qualitative behavior of the distribution such as the localization tendency.
At low and high $\mu$, the histograms are localized well both on the original and modified paths.
At $\mu = 1.25$ and $1.75$, on the other hand, the histograms are drastically changed by the path optimization, which leads to improvement of the signals.
In Eq.\,(\ref{eq:pq}), not only improvement in the denominator but also that in the numerator is important. 
As long as the mean value of APF is sufficiently larger than the statistical error, a well-localized distribution of the numerator shown in Fig.\,\ref{fig:scatter_plot_se_ori} can result in a small error of the physical observable.

\begin{figure}[t]
 \centering
 \includegraphics[keepaspectratio, scale=0.2]{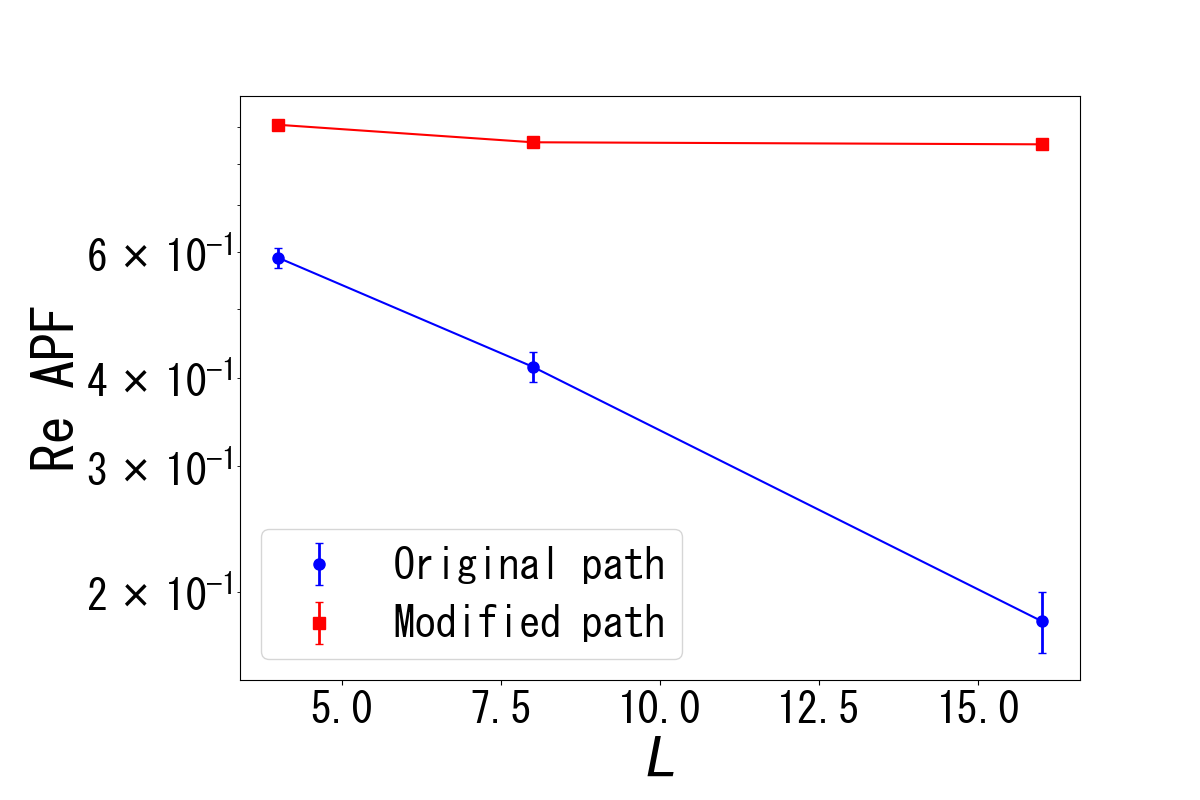}
 \includegraphics[keepaspectratio, scale=0.2]{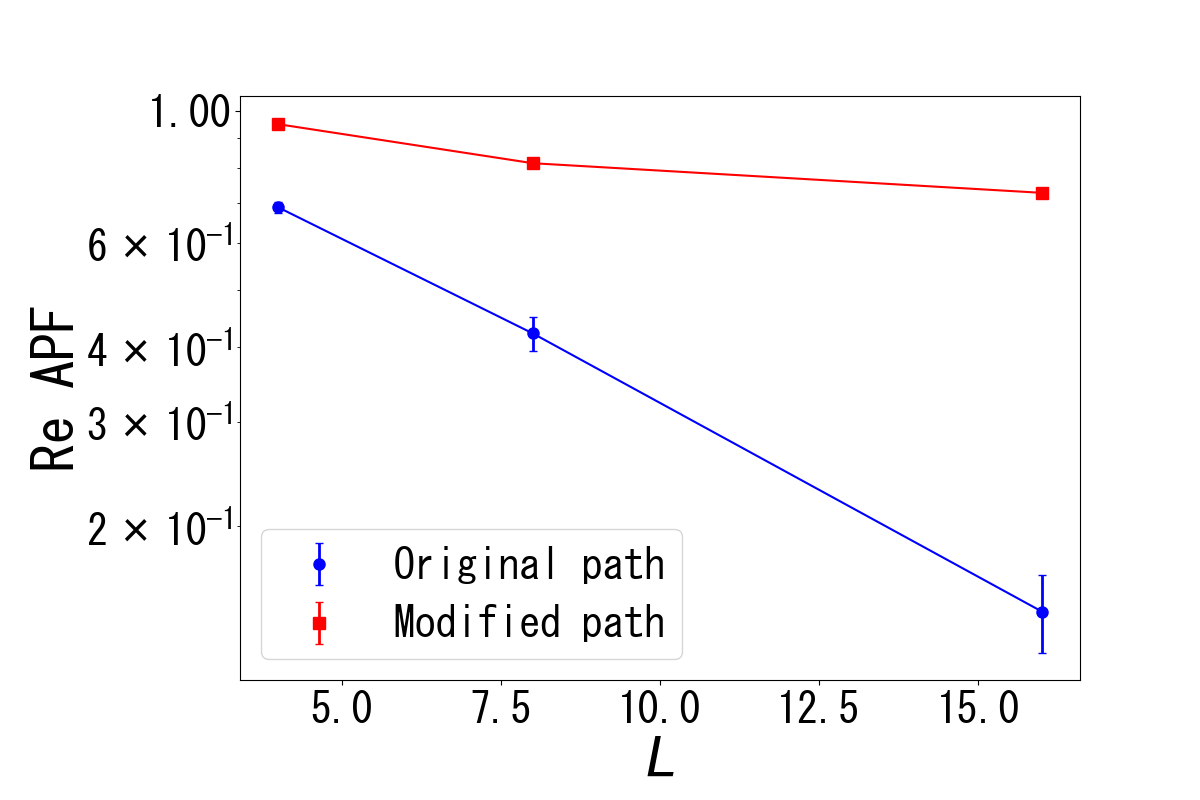}
 \caption{The upper (lower) panel shows the scaling behavior of APF with $\mu=1.0$ at $\beta=1$ ($\beta=2$).
 The circle and square symbols are the results on the original and modified integral paths, respectively.
 }
 \label{fig:scaling}
\end{figure}

The scaling behavior of APF at $\mu=1.0$ in terms of the system volume $L = 4, 8, 16$ is shown in Fig.\,\ref{fig:scaling}.
We see a similar scaling law in both cases,
\begin{align}
    \mathrm{APF} & \sim e^{- \alpha V},
\end{align}
see Ref.\,\cite{Splittorff:2007ck} as an example.
The effects of $\beta$ appear in $\alpha$; its effects are formulation dependent.
The modification of the integration path leads to a smaller value of $\alpha$, {\it i.e.,} the path optimization successfully improves the scaling behavior.
Of course, this method does not completely solve the curse of dimensionality.

\begin{figure}[t]
 \centering
 \includegraphics[keepaspectratio, scale=0.1]{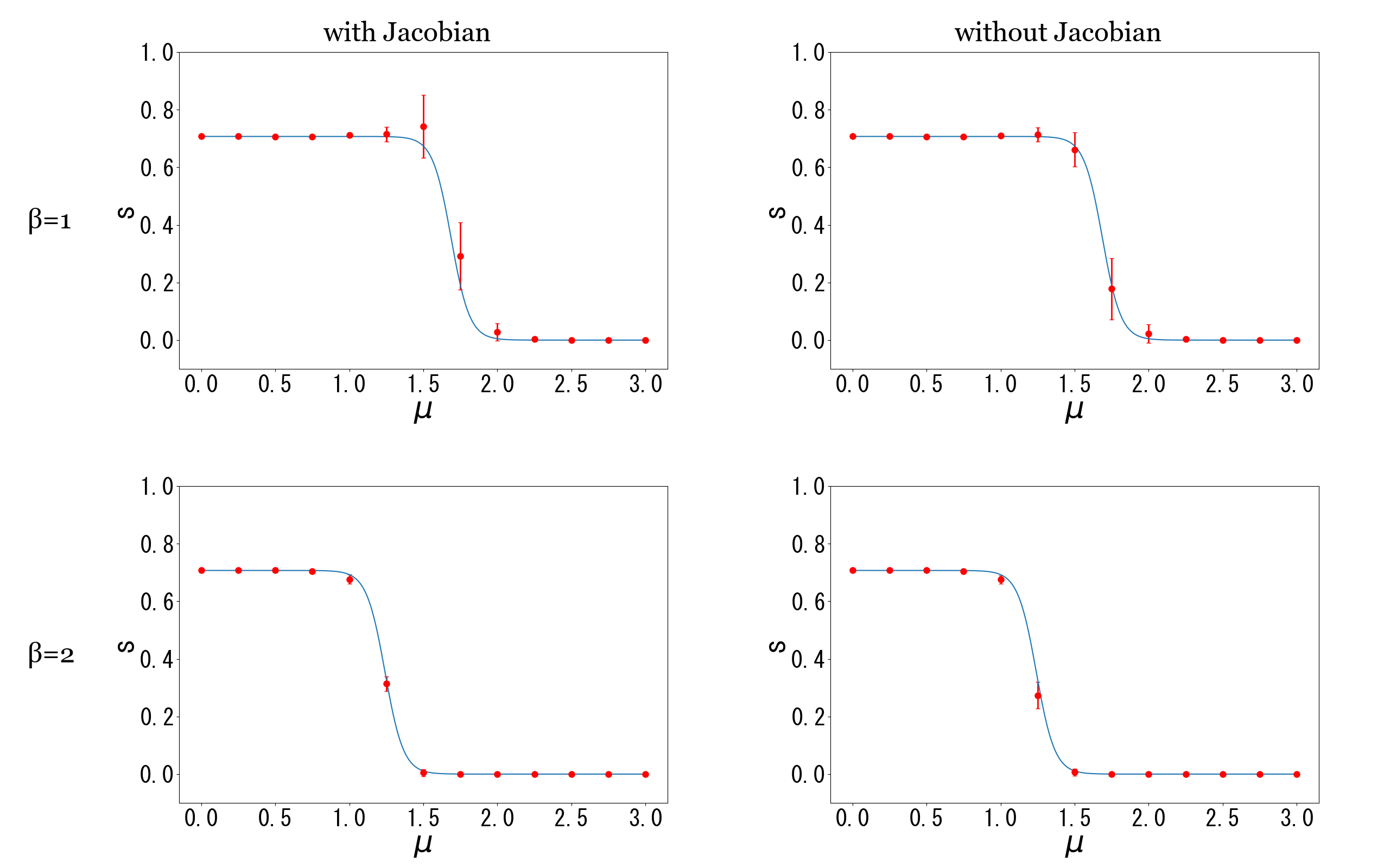}
 \caption{
 The $\mu$-dependence of the fermion condensate $s$ with $L = 16$.
 The symbols are our numerical results with the Jacobian calculation (left) and without the Jacobian calculation (right) in the learning process, and the lines denote the analytic results.
 The upper (lower) panels are the results at $\beta =1$ ($\beta=2$). }
 \label{fig:mu_s_j}
\end{figure}
\begin{figure}[t]
 \centering
 \includegraphics[keepaspectratio, scale=0.1]{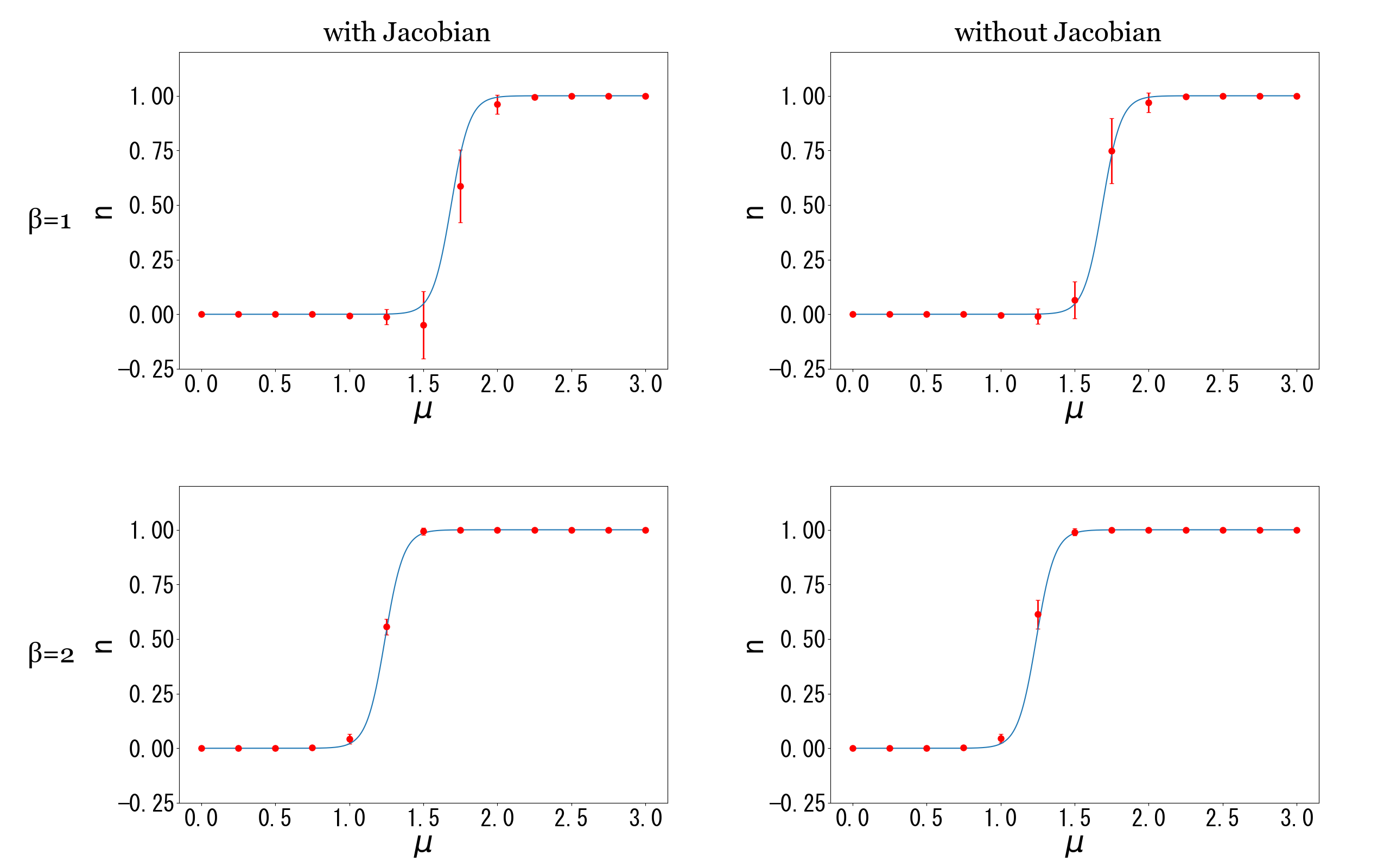}
 \caption{
 The $\mu$-dependence of the number density $n$ with $L = 16$.
 The symbols are our numerical results with the Jacobian calculation (left) and without the Jacobian calculation (right) in the learning process, and the lines denote the analytic results.
 The upper (lower) panels are results at $\beta =1$ ($\beta=2$).
 }
 \label{fig:mu_n_j}
\end{figure}

Next, we consider the approximation of the Jacobian in the learning step for the Thirring model.
The Jacobian calculation requires a large numerical cost ${\cal O}(N^3)$, where $N$ is the total degrees of freedom.
To reduce the cost, we employ the simplest approximation in which we replace the Jacobian matrix ${\cal J}$ with the unit matrix ${\mathbbm 1}$~\cite{Namekawa:2022liz};
\begin{align}
    J &= \mathrm{det} \, {\cal J} \to \mathrm{det} \, \mathbbm 1 = 1.
\end{align}
No numerical cost for the Jacobian calculation is required in the learning part.
The Jacobian calculation is required only in the evaluation step of the observables.
Figures~\ref{fig:mu_s_j} and \ref{fig:mu_n_j} show the $\mu$-dependence of the fermion condensate and the number density at $\beta =1$ and $2$ with and without the Jacobian calculation in the learning process.
In both cases of $\beta$, the approximation of the Jacobian works.
The results on the modified path reproduce the analytic values with comparable errors.
Since the lattice Thirring model is similar to QCD in terms of the origin of the sign problem, our result suggests that the simplest approximation of the Jacobian in the learning step also works in QCD.
It is noted that our simple Jacobian approximation indicates that the modification, such as bending and tilting, of integral path around the dominant points are not strong.
However, if the impacts of Jacobian on the calculation are not so weak, it is better to consider the following prescription: We first perform a few learning steps with the approximation of the Jacobian as a pre-training, and afterward, we perform the full learning.
It is expected to be an efficient procedure with significant cost reduction, especially in the complicated theory/model.
If this approximation cannot work, i.e., impacts of Jacobian are considerably strong, we must evaluate the Jacobian exactly.

\section{Summary}
\label{sec:summary}

In this paper, we have applied the path optimization method with machine learning to the one-dimensional massive lattice Thirring model as a laboratory to investigate the sign problem via the fermion-determinant term.
The modified integral path is represented by the neural network and the parameters are optimized via the self-supervise learning-like method.

We found that the path optimization method with machine learning works well in the Thirring model.
The average phase factor is enhanced on the modified integral path compared to the value on the original path, and our results agree with the analytic results with small statistical errors.
It indicates that the path optimization method works for the sign problem from the fermion-determinant term, which is of the same origin as that in QCD.

The approximation of Jacobian in the lattice Thirring model has also been examined.
We found that the perfect drop of the Jacobian calculation in the learning part, which significantly reduces the numerical cost, still works well.
The calculations with and without the Jacobian approximation give consistent expectation values of observables with small errors.

Based on this success, we apply the path optimization method to a more QCD-like theory/model, where the fermion determinant causes the sign problem.

\begin{acknowledgments}
The authors thank the late Prof. Akira Ohnishi for fruitful discussions at the early stage of this study.
This work is supported by the Japan Society for the Promotion of Science (JSPS) KAKENHI Grant Numbers (JP21K03553, 
JP22H05112, 
and JP24K07052
).
\end{acknowledgments}

\appendix

\section{Error estimation}
\label{sec:app_error}

\begin{figure}[t]
 \centering
 \includegraphics[keepaspectratio, scale=0.2]{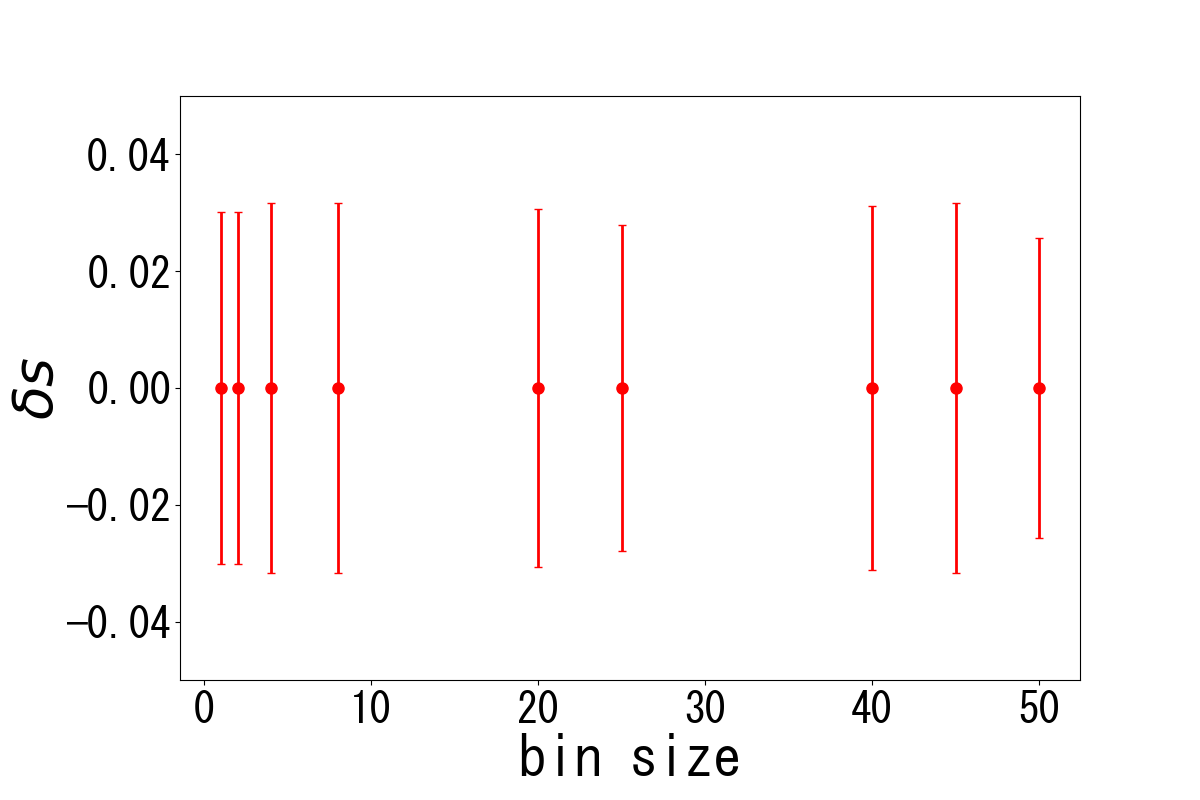}
 \caption{The bin size dependence of $\delta s$.
 }
 \label{fig:error}
\end{figure}

In this study, we estimate the statistic error by using the Jackknife method.
Figure \ref{fig:error} shows the statistic error of the fermion condensate $\delta s$ as a function of the bin size at $\beta=2$ and $\mu=1.25$, where the sign problem is severe in the model with the present setup.
Since we sample configurations from the Markov chain by separating $10$ times HMC update, the autocorrelation are quite small; the configurations are almost independent.
We set the bin size as $50$ in all computations to see the trend of the statistic error by using the path optimization method.

\newpage
\bibliography{ref.bib}

\end{document}